\documentclass[12pt]{iopart}
\usepackage{graphicx}
\usepackage{iopams}
\usepackage{dcolumn}
\usepackage{bm}
\usepackage{amsthm}
\usepackage{amssymb}
\usepackage[flushleft]{caption}
\usepackage{multirow}
\usepackage{subfigure}
\begin{document}

\title{Optimal condition for measurement observable via error-propagation}

\author{Wei Zhong$^1$, Xiao-Ming Lu$^2$, Xiao-Xing Jing$^1$ and Xiaoguang Wang$^1$}

\address{$^1$ Zhejiang Institute of Modern Physics, Department of Physics, Zhejiang University, Hangzhou 310027, China.}
\address{$^2$ Centre for Quantum Technologies, National University of Singapore, 3 Science Drive 2, Singapore 117543, Singapore.}
\ead{xgwang@zimp.zju.edu.cn}

\begin{abstract}
Propagation of error is a widely used estimation tool in experiments, where the estimation precision of the parameter depends on the fluctuation of the physical observable.
Thus which observable is chosen will greatly affect the estimation sensitivity.
Here we study the optimal observable for the ultimate sensitivity bounded by the quantum Cram\'er-Rao theorem in parameter estimation.
By invoking the Schr\"odinger-Robertson uncertainty relation, we derive the necessary and sufficient condition for the optimal observables saturating the ultimate sensitivity for single parameter estimate.
By applying this condition to Greenberg-Horne-Zeilinger states, we obtain the general expression of the optimal observable for separable measurements to achieve the Heisenberg-limit precision and show that it is closely related to the parity measurement.
However, Jose {\em et al} [Phys. Rev. A {\bf 87}, 022330 (2013)] have claimed that the Heisenberg limit may not be obtained via separable measurements.
We show this claim is incorrect.
\end{abstract}
\submitto{\JPA}
\maketitle

\section{Introduction}
An essential task in quantum parameter estimation is to suppress the fundamental bound on measurement precision imposed by quantum mechanics.
Various quantum strategies have been developed to enhance the accuracy of the parameter estimation, which are closely related to some practical applications, such as the Ramsey spectroscopies, atomic clocks, and the gravitational wave detection~\cite{Huelga1997,Chaves2013,Bollinger&Wineland1996,Buzek1999,LIGO2011,McKenzie2002,Vahlbruch2005,Kimble2001}.
Two approaches in common use for high-precision measurements are the parallel protocol with correlated multi-probes~\cite{Giovannetti&Lloyd2006} and multi-round protocol with a single probe~\cite{Higgins2007,Dam2007}.
Most recently, some novel methods, like environment-assisted metrology~\cite{Goldstein2011} and enhanced metrology by quantum error correction ~\cite{QEC1,QEC2,QEC3,QEC4}, were raised to achieve high precision in realistic experiments.

\begin{figure}[b]
    \includegraphics[width=14.8cm,height=3.4cm]{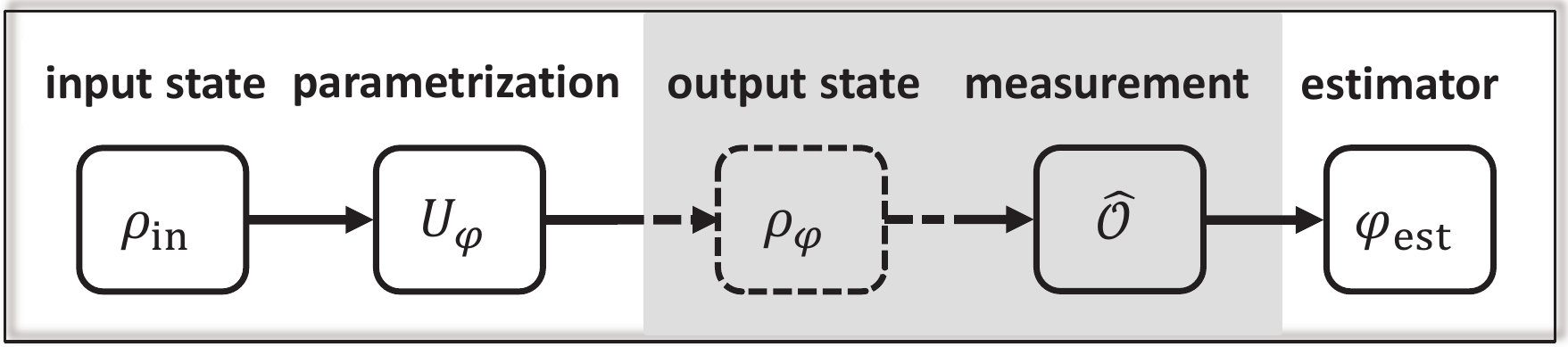}
     \caption{The schematic representation of a general scheme of (noiseless) quantum parameter estimation is composed of four components: input state $\rho_{\rm in}$, parametrization process $U_{\varphi}$, measurements $\hat{\mathcal{O}}$, and estimator $\varphi_{\rm est}$.
     Here, we concentrate on the part in shadow to find the optimal $\hat{\mathcal{O}}$ attaining the highest sensitivity to the parameter $\varphi$ in $\rho_{\varphi}$.}
    \label{fig:fig1}
\end{figure}

Rather than engineering the sensitivity-enhanced strategies, we concentrate on the problem of how to attain the maximal sensitivity in realistic experiments.
In general, a noiseless procedure of the quantum single parameter estimation can be abstractly modeled by four steps (see figure~\ref{fig:fig1}): (i) preparing the input state $\rho_{\rm in}$, (ii) parameterizing it under the evolution of the parameter-dependent Hamiltonian, for instance,  a unitary evolution $U_\varphi$ with $\varphi$ the parameter to be estimated, (iii) performing measurements of the observable $\hat{\mathcal{O}}$ on the output state $\rho_{\varphi}$, (iv) finally estimating the value of the parameter from the estimator $\varphi_{\rm est}$ as a function of the outcomes of the measurements.

From estimation theory, the estimation precision is statistically measured by the units-corrected mean-square deviation of the estimator $\varphi_{\rm est}$ from the true value $\varphi$ \cite{Braunstein&Caves1994,Braunstein&Caves1996},
\begin{equation}
    (\delta\varphi)^2_{\rm est} := \bigg\langle\bigg(\frac{\varphi_{\rm{est}}}{\vert \partial_\varphi\langle\varphi_{\rm{est}}\rangle_{\rm av}\vert}-\varphi\bigg)^2\bigg\rangle_{\rm av},
    \label{eq:uc_msd}
\end{equation}
where the brackets $\langle \, \rangle_{\rm av}$ denote statistical average and the derivative $\partial_\varphi\langle\varphi_{\rm{est}}\rangle\equiv\partial\langle\varphi_{\rm{est}}\rangle/\partial\varphi$ removes the local difference in the ``units" of $\varphi_{\rm{est}}$ and $\varphi$.
Whatever is the measurement scheme employed, the ultimate limit to the precision of the unbiased estimate is given by the quantum Cram\'er-Rao bound (QCRB) from below as
\begin{equation}
    (\delta\varphi)^2_{\rm est} \geq (\upsilon\mathcal{F}_{\varphi})^{-1},
    \label{eq:QCRB}
\end{equation}
where $\upsilon$ is the repetitions of the experiment and $\mathcal{F}_{\varphi}$ is the quantum Fisher information (QFI) (see equation~(\ref{eq:QFI}) for definition), which measures the statistical distinguishability of the parameter in quantum states.
This bound is asymptotically achieved for large $\upsilon$ under optimal measurements followed by the maximum likelihood estimator~\cite{Braunstein&Caves1994,Braunstein&Caves1996,HelstromBook,HolevoBook}.

On the other hand, it is well-known that error-propagation is a widely acceptable theory in experiments~\cite{Huelga1997,Bollinger&Wineland1996,Giovannetti&Lloyd2006,Yurke1986,Kim1998,Campos2003,Hotta&Ozawa2004,Durkin&Dowling2007,Pezze&Smerzi2008,Boixo&Caves2008PRL,Boixo&Caves2008PRA,Anisimov&Dowling2010,Gerry&Mimith2010,Hofmann2009,Joo2011,Seshadreesan2013}.
With this theory, to estimate the parameter $\varphi$ is reduced to measuring the average value of a physical observable $\hat{\mathcal{O}}$.
After repeating the experiment $\upsilon$ times, the real accessible precision on $\varphi$ is given by the error-propagation formula as follows \cite{Huelga1997,Bollinger&Wineland1996,Giovannetti&Lloyd2006,Yurke1986,Kim1998,Campos2003,Hotta&Ozawa2004},
\begin{equation}
    (\delta\varphi)^2_{\rm ep}
    := \frac{1}{\upsilon}\frac{\langle(\Delta \hat{\mathcal{O}})^{2}\rangle}{\,\,\vert\partial_{\varphi}\langle \hat{\mathcal{O}}\rangle\vert^{2}},
    \label{eq:error-propagation}
\end{equation}
where $\Delta \hat{\mathcal{O}} = \hat{\mathcal{O}} - \langle \hat{\mathcal{O}} \rangle$ and $\langle \hat{\mathcal{O}} \rangle = {\rm Tr}(\rho_{\varphi} \hat{\mathcal{O}})$.
Note that the two estimation errors defined in equations (\ref{eq:uc_msd}) and (\ref{eq:error-propagation}) are closely related.

To show the relationship between the two kinds of the estimation errors, $(\delta\varphi)^2_\mathrm{est}$ and $(\delta\varphi)^2_\mathrm{ep}$, we introduce $\Delta\varphi_\mathrm{est}:=\varphi_\mathrm{est}-\langle \varphi_\mathrm{est}\rangle_\mathrm{av}$. Then, it is easy to show that~\cite{Braunstein&Caves1994}
\begin{equation}
(\delta\varphi)^2_\mathrm{est}=
\frac{\langle (\Delta\varphi_{\rm est})^2\rangle_{\rm av}}{|\partial_\varphi\langle\varphi_{\rm{est}}\rangle_{\rm av}|^2}
+\bigg\langle\frac{\varphi_{\rm{est}}}{|\partial_\varphi\langle\varphi_{\rm{est}}\rangle_{\rm av}|}-\varphi\bigg\rangle_{\rm av}^2.
\end{equation}
When viewing the arithmetic mean of the measurement outcomes of $\hat{\mathcal{O}}$ over repetitions of the experiment as the estimator in the quantum setting,
one has in general $(\delta\varphi)^2_{\rm est}\geq(\delta\varphi)^2_{\rm ep}\geq(\upsilon\mathcal{F}_\varphi)^{-1}$ by noting that $\langle(\Delta\varphi_{\rm est})^2\rangle=\langle (\Delta\hat\mathcal{O})^2\rangle/\upsilon$ for sufficiently large $\upsilon$ according to the central limit theorem~\cite{central_limit} and comparison of the two definitions of the errors given by equations~(\ref{eq:uc_msd}) and (\ref{eq:error-propagation}).
In such situation, $(\delta\varphi)^2_{\rm est}$ and $(\delta\varphi)^2_{\rm ep}$ have the same QCRB, and the saturation of the former implies that of the latter.

The formula equation~(\ref{eq:error-propagation}) indicates that the fluctuation of the observable $\hat{\mathcal{O}}$ propagates to the estimated values of the parameter $\varphi$.
This means that what kinds observable $\hat{\mathcal{O}}$ employed directly affects the estimating precision of the parameter $\varphi$.
The purpose of this paper is to address the question of with which kind of observable does the estimation error given by equation~(\ref{eq:error-propagation}) achieve the QCRB given by equation~(\ref{eq:QCRB}).

In this paper, we derive the necessary and sufficient (N\&S) condition for the optimal observable saturating the QCRB for the single parameter estimation by using the Schr\"odinger-Robertson uncertainty relation (SRUR).
We then apply this condition to GHZ states and find the general form of the optimal observable for separable measurements to achieve the Heisenberg-limit sensitivity (i.e., $1/N$).
Moreover, we discuss the relation between the optimal separable observable and parity measurements.
However, Jose {\em et al.}, in a recent work \cite{Jose&Shaji2013}, made a contradictory conclusion with respect to the above result.
They claimed that separable measurements are impossible to go beyond the shot-noise limit (i.e., $1/{\sqrt{N}}$) for any entangled states.
To clarify this issue, we revisit the method in \cite{Jose&Shaji2013} and show the causes for this inconsistency.

This paper is structured as follows.
In section~\ref{sec:NS}, we first briefly review the single parameter estimation and obtain the N\&S condition for the optimal observable.
In section~\ref{sec:example}, we give an application of this condition to obtain the optimal separable observables for GHZ states to saturate the Heisenberg-limit precision.
In section~\ref{sec:discussion}, we further elucidate the reasons for contradiction between the result given in \cite{Jose&Shaji2013} and ours.
At last, a conclusion is given in section~\ref{sec:conclusion}.

\section{N\&S condition for optimal observable in single parameter estimation\label{sec:NS}}

We start by a brief review of quantum single parameter estimation via the general estimator.
Consider a parametric family of density matrices $\rho_\varphi$ containing an unknown parameter $\varphi$ to be estimated.
Suppose that the general quantum measurement performed on $\rho_\varphi$ is characterized by a positive-operator-valued measure $\hat{M} := \{\hat{M}_x\}$ with $x$ the results of measurement.
The value of the parameter is inferred via an estimator $\varphi_{\rm est}$, which maps the measurement outcomes to the estimated value.
After repeating the experiment $\upsilon$ times, the standard estimation error $(\delta \varphi)^2_{\rm est}$ in equation~(\ref{eq:uc_msd}) is bounded from below as
\begin{equation}
    (\delta \varphi)^2_{\rm est}
    \geq (\upsilon F_\varphi)^{-1}, \label{eq:CRB}
\end{equation}
where
\begin{equation}
    F_\varphi := \sum_x p_\varphi(x) [\partial_\varphi \ln p_\varphi(x)]^2
\end{equation}
is the (classical) Fisher information of the measurement-induced probability distribution $p_\varphi(x)=\mathrm{Tr}(\rho_\varphi \hat{M}_x)$.
The maximization over all POVMs gives rise to the so-called QFI, which is defined by
\begin{equation}
    \mathcal{F}_{\varphi} := {\rm Tr}(\rho_{\varphi} \hat{L}_\varphi^{2}).
    \label{eq:QFI}
\end{equation}
Hence, a more tighter bound of equation~(\ref{eq:CRB}) is given by equation~(\ref{eq:QCRB}).
Here $\hat{L}_\varphi$ is the symmetric logarithmic derivative (SLD) operator, which is a Hermitian operator determined by
\begin{equation}
    \partial_{\varphi}\rho_{\varphi}=\frac{1}{2}[\rho_{\varphi}, \hat{L}_\varphi]_{+}
    \label{eq:SLD}
\end{equation}
with $[\cdot\,,\,\cdot]_{+}$ denoting the anti-commutator, see reference~\cite{Braunstein&Caves1994}.
It is remarkable that $\hat{L}_\varphi$ may not be uniquely determined by equation~(\ref{eq:SLD}) when $\rho_{\varphi}$ is not of full rank \cite{Fujiwara1999}.

However, in general the value of the parameter $\varphi$ may not be directly measured.
The most general method of estimating the value of $\varphi$ in practice involves measurements corresponding to a physical observable $\hat{\mathcal{O}}$ which is generally $\varphi$-independent.
In such cases, the estimation error is given by the error-propagation formula equation~(\ref{eq:error-propagation}), in which the fluctuations on the observable $\hat{\mathcal{O}}$ propagate to the uncertainty in the estimation of $\varphi$.
In the following, we follow Hotta and Ozawa \cite{Hotta&Ozawa2004} to derive the achievable lower bound of the estimation error $(\delta\varphi)^2_{\rm ep}$ by using the SRUR.

Let us first recall the SRUR \cite{SchrodingerUnc,RobertsonUnc}, which states that the uncertainty of two non-commuting observables $\hat{X},\,\hat{Y}$ must obey the following inequality
\begin{equation}
    \langle (\Delta \hat{X})^{2}\rangle \langle (\Delta \hat{Y})^{2}\rangle
    \geq \frac{1}{4}\vert\langle [\hat{X}, \hat{Y}]\rangle\vert^{2}+\frac{1}{4}\langle[\Delta \hat{X},\Delta \hat{Y}]_{+}\rangle^{2},
    \label{eq:SRUR}
\end{equation}
where $[\cdot\,,\,\cdot]$ denotes the commutator.
The SRUR follows from the Schwarz inequality for the Hilbert-Schmidt inner product, and naturally reduces to the Heisenberg uncertainty relation  under the condition $\langle[\Delta \hat{X},\Delta \hat{Y}]_{+}\rangle =0$.
By substituting $\hat{X}\,(\hat{Y})$ with $\hat{\mathcal{O}}\,(\hat{L}_\varphi$) and utilizing
\begin{equation}
    \mathcal{F}_{\varphi}=\langle \hat{L}_\varphi^{2}\rangle =\langle (\Delta \hat{L}_\varphi)^{2}\rangle,
    \label{eq:QFI-1}
\end{equation}
as a result of $\langle \hat{L}_\varphi \rangle = 2\,\partial_\theta \mathrm{Tr}(\rho_\varphi) = 0$, equation~(\ref{eq:SRUR}) becomes
\begin{equation}
    \langle (\Delta \hat{\mathcal{O}})^{2}\rangle \,\mathcal{F}_{\varphi}
    \geq
    \frac{1}{4}\vert\langle[\hat{\mathcal{O}}, \hat{L}_\varphi]\rangle\vert^{2}+\frac{1}{4}\langle [\hat{\mathcal{O}}, \hat{L}_\varphi]_{+}\rangle^{2}.\label{eq:SRUR_1}
\end{equation}
Moreover, since the observable operator $\hat{\mathcal{O}}$ is independent of $\varphi$, we have
\begin{eqnarray}
    \langle [\hat{\mathcal{O}}, \hat{L}_\varphi]_+ \rangle &=& \mathrm{Tr}([\hat{\mathcal{O}}, \hat{L}_\varphi]_+\rho_\varphi) \nonumber\\
    &=& \mathrm{Tr}(\hat{\mathcal{O}}[\hat{L}_\varphi, \rho_\varphi]_+) \nonumber\\
    &=& 2\,\partial_\varphi\langle  \hat{\mathcal{O}} \rangle,
    \label{eq:equalities}
\end{eqnarray}
where the second equality is obtained by employing the cyclic property of the trace operation, and the third equality is due to  the SLD equation (\ref{eq:SLD}).
Provided that $\langle \hat{\mathcal{O}} \rangle$ is nonzero, combining equations~(\ref{eq:error-propagation}), (\ref{eq:SRUR_1}) and (\ref{eq:equalities}) yields
\begin{eqnarray}
    (\delta\varphi)^2_{\rm ep}
    & \geq & \frac{1}{\upsilon\mathcal{F}_{\varphi}}\bigg(1+\frac{\vert\langle[\hat{\mathcal{O}}, \hat{L}_\varphi]\rangle\vert^{2}}{\langle [\hat{\mathcal{O}}, \hat{L}_\varphi]_{+}\rangle^{2}}\bigg)
    \label{eq:GQCRB}\\
    & = & \frac{1}{\upsilon\mathcal{F}_{\varphi}}\bigg[1+\bigg(\frac{{\rm Im}\langle \hat{\mathcal{O}}\hat{L}_\varphi\rangle}{{\rm Re}\langle \hat{\mathcal{O}}\hat{L}_\varphi\rangle }\bigg)^{2}\bigg] \\
    &\geq& (\upsilon\mathcal{F}_{\varphi})^{-1}.
    \label{eq:CR_bound}
\end{eqnarray}
The bound in equation~(\ref{eq:GQCRB}) describes the achievable sensitivity of $\varphi$ when employing an observable $\hat{\mathcal{O}}$.
The bound in equation~(\ref{eq:CR_bound}) gives the highest precision for $\varphi$ for the optimal observable $\hat{\mathcal{O}}_{\rm opt}$, which coincides with the QCRB in equation~(\ref{eq:QCRB}).
It is shown that the estimation error $(\delta\varphi)^2_{\rm ep}$ achieves the QCRB only when the two equalities in equations (\ref{eq:GQCRB}) and (\ref{eq:CR_bound}) hold simultaneously.

Below, we consider the attainability of the above bounds and give the N\&S condition for optimal observables.
From the N\&S condition for equality in the SRUR, the equality in equation~(\ref{eq:GQCRB}) holds if and only if
\begin{equation}
    \Delta \hat{\mathcal{O}} \sqrt{\rho_\varphi} = \alpha \hat{L}_\varphi \sqrt{\rho_\varphi}
    \label{eq:condition1}
\end{equation}
is satisfied with a nonzero complex number $\alpha$.
Note that we restrict here $\alpha\neq0$ at the request of $\langle [\hat{\mathcal{O}}, \hat{L}_\varphi]_{+}\rangle\neq0$  in the denominator of equation~(\ref{eq:GQCRB}).
Furthermore, the equality in equation~(\ref{eq:CR_bound}) holds if and only if
\begin{equation}
    \mathrm{Im} \langle \hat{\mathcal{O}} \hat{L}_\varphi \rangle = 0.
    \label{eq:condition2}
\end{equation}
This condition can be combined into the condition~(\ref{eq:condition1}) by restricting $\alpha$ to be a nonzero real number, i.e.,
\begin{equation}
    \Delta \hat{\mathcal{O}} \sqrt{\rho_\varphi} = \alpha \hat{L}_\varphi \sqrt{\rho_\varphi} \quad \mbox{with $\alpha \in \mathbb{R} \!\setminus\!\! \{0\}$}.
    \label{eq:NS_condition}
\end{equation}
This is the of the optimal observable for density matrix $\rho_\varphi$.
It implies that the estimation error achieves the QCRB given by the QFI for $\rho_\varphi$ {\em only} when the observable that we choose satisfies equation~(\ref{eq:NS_condition}).
This is the main result of the paper.
For pure states $\rho_\varphi = \vert\psi_\varphi\rangle \langle\psi_\varphi\vert$, the condition~(\ref{eq:NS_condition}) is equivalent to
\begin{equation}
    \Delta \hat{\mathcal{O}}\vert\psi_\varphi\rangle = \alpha \hat{L}_\varphi\vert\psi_\varphi\rangle \quad \mbox{with $\alpha \in \mathbb{R} \!\setminus\!\! \{0\}$}.
    \label{eq:cond_pure}
\end{equation}

If we assume that the parameter $\varphi$ here is imprinted via a unitary operation \cite{Giovannetti&Lloyd2006}, i.e., $\rho_{\varphi}=\exp({-i\hat{G}\varphi})\,\rho_{\rm in}\exp({i\hat{G}\varphi})$ with $\hat{G}$ the generator, associating with the equality $\partial_{\varphi}\rho_{\varphi}=-i[\hat{G},\rho_{\varphi}]$, then condition~(\ref{eq:cond_pure}) further reduces to
\begin{equation}
    \Delta \hat{\mathcal{O}}\vert\psi_{\varphi}\rangle =
    -2i\alpha\Delta \hat{G}\vert\psi_{\varphi}\rangle \quad \mbox{with $\alpha \in \mathbb{R} \!\setminus\!\! \{0\}$}.
    \label{eq:cond_pure_Hofmann}
\end{equation}
This condition was alternatively obtained in Ref.~\cite{Hofmann2009}.
It is deserved to note that their proof is only valid in the case of unitary parametrization for pure states, and cannot be generalized to obtain the condition (\ref{eq:NS_condition}).

Here, we discuss the relations between the saturation of the QCRB with respect to $(\delta\varphi)^2_{\rm est}$ and that with respect to $(\delta\varphi)^2_{\rm ep}$.
Following Braunstein and Caves~\cite{Braunstein&Caves1994}, the saturation of the QCRB with respect to the error $(\delta\varphi)^2_{\rm est}$ can be separated as the saturation of a classical Cram\'er-Rao bound (CCRB) equation~(\ref{eq:CRB}) and finding an optimal measurement attaining the QFI.
The CCRB can always be asymptotically achieved by the maximum likelihood estimator, so whether the QCRB can be asymptotically saturated is determined by whether the measurement attains the QFI.
The  N\&S condition for the optimal measurement attaining the QFI reads~\cite{Braunstein&Caves1994}
\begin{equation}\label{eq:optimal_POVM}
	\sqrt{\hat{M}_x}\sqrt{\rho_\varphi} = u_x \sqrt{\hat{M}_x} \hat{L}_\varphi \sqrt{\rho_\varphi},
\end{equation}
where $\{\hat M_x\}$ denotes the POVM of the measurement and $u_x$ are real numbers.
In the following, we show that the N\&S condition (\ref{eq:NS_condition}) for the saturation of the QCRB with respect to $(\delta\varphi)^2_{\rm ep}$ identifies an optimal measurement attaining QFI.
Let $\hat\mathcal O_{\rm opt}$ be the optimal observable satisfying Eq.~(\ref{eq:NS_condition}) and $P_x$ the eigenprojectors of $\hat\mathcal O_{\rm opt}$ with the eigenvalues $x$. Left multiplying $P_x$ on both sides of Eq.~(\ref{eq:NS_condition}), it is easy to see that $\{P_x\}$ is the optimal measurement attaining the QFI.
That is to say, the projective measurement $\{P_x\}$ followed by the maximum likelihood estimator of the measurement outcomes saturate the QCRB with respect to the standard estimation error $(\delta\varphi)^2_{\rm est}$.

\section{Optimal separable observable for GHZ states \label{sec:example}}
Below, we apply the N\&S condition to show the general optimal observable for GHZ states.
Let us specifically consider an experimentally realizable Ramsey interferometry to estimate the transition frequency $\omega$ of the two-level atoms loaded in the ion trap \cite{Huelga1997,Chaves2013}.
The Hamiltonian of the system with $N$ atoms is $\hat{H} = (\omega/2)\sum_{i=1}^{N}\hat{\sigma}_z^i$ where $\hat{\sigma}_z^i$ is the Pauli matrix acting on the $i$th particle.
In this setup, the measurements are limited to be performed separately on each atom.
The observable operator may be described as a tensor product of Hermitian matrices $\hat{\mathcal{O}} = \hat{\mathcal{O}}_{\rm q}^{\otimes N}$ with $\hat{\mathcal{O}}_{\rm q}=a_0 \mathbb{I}+\bm{a}\cdot\hat{\bm{\sigma}}$ dependent of four real coefficients $\{a_0,a_1,a_2,a_3\}$, where $\mathbb{I}$ is the identity matrix of dimension $2$.

Suppose that the input state is the maximally entangled states, i.e., GHZ states, which provides the Heisenberg-limit-scaling sensitivity of frequency estimation in the absence of noise \cite{Huelga1997,Giovannetti&Lloyd2006,GHZ1989Book}.
Under the time evolution $\hat{U} = \exp{(-i\hat{H}t)}$, the output state can be represented as
\begin{equation}
    \vert\psi_{\rm GHZ}(\varphi)\rangle 	
    = \frac{1}{\sqrt{2}}\big(\vert 0\rangle^{\otimes N} +e^{iN\varphi}\vert1\rangle^{\otimes N}\big),
    \label{eq:GHZ_varphi}
\end{equation}
up to an irrelevant global phase with $\varphi=\omega t$.
Here, we adopt the standard notation where $\vert 0\rangle$ and $\vert 1\rangle$ are the eigenvectors of $\sigma_z$ corresponding to eigenvalues $+1$ and $-1$, respectively.
To determine the optimal separable observable $\hat{\mathcal{O}}$, we need to find the solutions of the coefficients $\{a_0,a_1,a_2,a_3\}$ to satisfy equation~(\ref{eq:cond_pure}).
With $\hat{L}_\varphi=2\partial_\varphi(\vert\psi_\varphi\rangle\langle\psi_\varphi\vert)$ for pure states, the SLD operator for the state of equation~(\ref{eq:GHZ_varphi}) is given by
\begin{equation}
    \hat{L}_\varphi = -iNe^{-iN\varphi}\left(\vert0\rangle\langle1\vert\right)^{\otimes N}
    +iNe^{iN\varphi}\left(\vert1\rangle\langle0\vert\right)^{\otimes N}.
    \label{eq:SLD_GHZ_varphi}
\end{equation}
We find that equation~(\ref{eq:cond_pure}) is always satisfied for $a_0=a_3=0$ and arbitrary real number $a_1,\,a_2$ that do not vanish simultaneously.
Therefore, the general expression of the optimal separable observable is given by
\begin{equation}
    \hat{\mathcal{O}}_{\rm opt} = (a_{1}\hat{\sigma}_{x}+a_{2}\hat{\sigma}_{y})^{\otimes N},
    \label{eq:A_separable_GHZ}
\end{equation}
which is independent of the parameter $\varphi$, i.e., globally optimal in the whole range of the parameter.
It is easy to check that such observables saturate the Heisenberg-limit sensitivity.
Actually, according to the error-propagation formula equation~(\ref{eq:error-propagation}), we have
\begin{equation}
    \delta\varphi_{\rm GHZ}
    = \frac{1}{\sqrt{\upsilon}}\frac{\sqrt{\langle \hat{\mathcal{O}}_{\rm opt}^{2}\rangle-\langle \hat{\mathcal{O}}_{\rm opt}\rangle^2}}
    {\vert\partial_{\varphi}\langle \hat{\mathcal{O}}_{\rm opt}\rangle\vert}
    = \frac{1}{\sqrt{\upsilon}N},
    \label{eq:error-propagation_GHZ}
\end{equation}
as a result of
\begin{eqnarray}
    \langle \hat{\mathcal{O}}_{\rm opt}\rangle &=& {\rm Re}[e^{-iN\varphi}(a_{1}+ia_{2})^{N}],
    \label{eq:expectation1}\\
    \langle \hat{\mathcal{O}}_{\rm opt}^{2}\rangle &=& (a_{1}^{2}+a_{2}^{2})^{N}.
    \label{eq:expectation2}
\end{eqnarray}
When setting $a_1=1, a_0=a_2=a_3=0$, the optimal observable in equation~(\ref{eq:A_separable_GHZ}) reduces to $\hat{\sigma}_{x}^{\otimes N}$, as given in~\cite{Giovannetti&Lloyd2006}.
Note that here measuring the observable $\hat{\sigma}_{x}^{\otimes N}$ fails to attain the Heisenberg limit for the cases of $\varphi=k\pi/N,\,(k\in\mathbb{Z})$ in which equation~(\ref{eq:error-propagation_GHZ}) becomes singular.
Besides, we note that measuring the spin observable $\hat{\sigma}_{y}^{\otimes N}$ also fail in these cases when $N$ is even, and it is useful except for the cases of $\varphi=(2k+1)\pi/2N,\,(k\in\mathbb{Z})$ when $N$ is odd.

We next show that the optimal observable in the form of equation~(\ref{eq:A_separable_GHZ}) is closely related to the parity measurement proposed originally by Bollinger {\em et al} \cite{Bollinger&Wineland1996}.
As is well known, in the standard Ramsey interferometry, there are generally two Ramsey pulses applying before and after the free evolution (with an accumulated phase $\varphi$), and measurements often take place after the second pulse \cite{Huelga1997,Bollinger&Wineland1996}.
Here the action of the pulse is modeled by a $\pi/2$-rotation operation about the $y$ axis, i.e., $R_y\big[\frac{\pi}{2}\big]=\exp[-i(\frac{\pi}{2})\hat{J}_y]$, and the measurement observable is denoted as the operator $\hat{\mathcal{O}}_f$.
With equation~(\ref{eq:A_separable_GHZ}), one has
\begin{equation}
    \hat{\mathcal{O}}_f = R_y^\dagger\bigg[\frac{\pi}{2}\bigg]\,\hat{\mathcal{O}}\,R_y\bigg[\frac{\pi}{2}\bigg] =
     (a_{1}\hat{\sigma}_{z}+a_{2}\hat{\sigma}_{y})^{\otimes N}.
     \label{eq:A_separable_GHZ_final}
\end{equation}
When setting $a_1=1, a_2=0$, equation~(\ref{eq:A_separable_GHZ_final}) reduces to
\begin{equation}
\hat{\mathcal{O}}_f = \hat{\sigma}_{z}^{\otimes N} \equiv (-1)^{j-\hat{J}_z}
\end{equation}
with $j=N/2$, which is the so-called parity measurement \cite{Bollinger&Wineland1996}.
It is shown that only a parity measurement is necessary for the optimal estimate of the phase parameter $\varphi$ for GHZ states, and it is more experimentally feasible than the detection strategy, as discussed in~\cite{Giovannetti&Lloyd2006}, that applies local operations and classical communication.

\section{Further discussions \label{sec:discussion}}
However, in a recent work \cite{Jose&Shaji2013}, it was pointed out that the separable measurement (the restricted readout procedure) might not be possible to go beyond the shot-noise limit even for arbitrary entangled states.
It seems that this conclusion is inconsistent with ours in the above discussion.
In what follows, we clarify this issue by revisiting the method in~\cite{Jose&Shaji2013} and showing the causes for this inconsistency.

For simplicity, let us consider the two-qubit parametric GHZ state
\begin{eqnarray}
    \vert\psi^{(2)}_{\rm GHZ}(\varphi)\rangle
    = \frac{1}{\sqrt{2}}\big(\vert 00\rangle +e^{2i\varphi}\vert 11\rangle\big).
    \label{eq:GHZ_varphi_2}
\end{eqnarray}
Following Ref.~\cite{Jose&Shaji2013}, we restrict the separable measurement to be the projective measurements $\{\vert+\rangle \langle+\vert, \vert-\rangle \langle-\vert\}$ for each qubit with
\begin{equation}
    \vert\pm\rangle = \frac{1}{\sqrt{2}}(\vert 0\rangle\pm\vert 1\rangle).
    \label{eq:local_measurement}
\end{equation}
According to the condition of equation~(\ref{eq:optimal_POVM}), whether the above restricted measurement presented by equation~(\ref{eq:local_measurement}) is the optimal measurement saturating the QCRB can be tested by asking whether or not the operators of the form
\begin{eqnarray}
    \hat{K} &=& \lambda_{++}\vert++\rangle\langle++\vert
    +\lambda_{+-}\vert+-\rangle\langle+-\vert \nonumber\\
    & & +\,\lambda_{-+}\vert-+\rangle\langle-+\vert
    +\lambda_{--}\vert--\rangle\langle--\vert
    \label{eq:SLD_projective_2}
\end{eqnarray}
can be the SLD operator for the state of equation~(\ref{eq:GHZ_varphi_2}).
By domenstrating that for the state in equation~(\ref{eq:GHZ_varphi_2}) with $\varphi=0$, there is no solution of the SLD equation~(\ref{eq:SLD}) for the coefficients $\{\lambda_{++},\lambda_{+-},\lambda_{-+},\lambda_{--}\}$ in equation~(\ref{eq:SLD_projective_2}), the authors in Ref.~\cite{Jose&Shaji2013} claimed that the projective measurement about $\{\vert++\rangle, \vert+-\rangle, \vert-+\rangle, \vert--\rangle\}$ is not the optimal measurement for the state of equation~(\ref{eq:GHZ_varphi_2}).

However, as we showed in the Sec. 3, $\sigma_x\otimes\sigma_x$ is an optimal observable saturating the QCRB with respect to $(\delta\varphi)^2_{\rm ep}$ for the states (\ref{eq:GHZ_varphi_2}).
Although the estimation error considered in the Ref.~\cite{Jose&Shaji2013} is $(\delta\varphi)^2_{\rm est}$, a contradiction still arises, as the projective measurement of $\sigma_x\otimes\sigma_x$ attains the QFI of states (\ref{eq:GHZ_varphi_2}) (see the end in Sec. 2) so that $\{\vert++\rangle, \vert+-\rangle, \vert-+\rangle, \vert--\rangle\}$ is the optimal measurement regarding the estimation error $(\delta\varphi)^2_{\rm est}$.
Below, we shall show that actually for any other point except for $\varphi=k\pi/2,\,(k\in\mathbb{Z})$ in the range of the parameter, there do exist the SLD operator in form of equations~(\ref{eq:SLD_projective_2}).

First, note that the SLD operator for the non-full-rank density matrices is not uniquely determined, but $\hat{L}_\varphi\rho_\varphi$ (or $\hat{L}_\varphi\vert\psi_\varphi\rangle$ for pure state) is uniquely determined.
Second, from equation~(\ref{eq:SLD_GHZ_varphi}), we see
\begin{equation}
    \hat{L}_\varphi = -2ie^{-2i\varphi}\vert00\rangle\langle11\vert + 2ie^{2i\varphi}\vert11\rangle\langle00\vert.
    \label{eq:SLD_GHZ}
\end{equation}
is a SLD operator for the state of equation~(\ref{eq:GHZ_varphi_2}).
Third, since $\hat{L}_\varphi\vert\psi_\varphi\rangle$ is uniquely determined, then if $\hat{K}$ is the SLD operator for $\vert\psi_\varphi\rangle$ if and only if
\begin{equation}
    \hat{L}_\varphi\vert\psi_\varphi\rangle = \hat{K}\vert\psi_\varphi\rangle
    \label{eq:SLD_pure_eq}
\end{equation}
is satisfied.
Thus, substituting equations~(\ref{eq:GHZ_varphi_2}), (\ref{eq:SLD_projective_2}) and (\ref{eq:SLD_GHZ}) into equation~(\ref{eq:SLD_pure_eq}), we obtain the solutions for the coefficients as
\begin{equation}
    \lambda_{++} = \lambda_{--} = -2\tan\varphi,\quad \lambda_{+-} = \lambda_{-+} = 2\cot\varphi.
    \label{eq:shaji_solution}
\end{equation}
The above solutions are singular for $\varphi=k\pi/2,\,(k\in\mathbb{Z})$, which coincide with the results discussed below equation~(\ref{eq:expectation2}).
Note that here the $\varphi=0,\,(k=0)$ case is just considered in Ref.~\cite{Jose&Shaji2013}.
Whilst, for a general value of the parameter except those singular points, the restricted separable measurement considered here indeed saturate the Heisenberg-limit-scaling sensitivity for the parametric state of equation~(\ref{eq:GHZ_varphi_2}).
Moreover, it is easy to check that the same results of Eq.~(\ref{eq:shaji_solution}) can be obtained when restricting the separable measurement to be the projective measurements $\{\vert+\rangle_y \langle+\vert, \vert-\rangle_y \langle-\vert\}$ for each qubit with
\begin{equation}
    \vert\pm\rangle_y = \frac{1}{\sqrt{2}}(\vert 0\rangle\pm i\vert 1\rangle)
    \label{eq:local_measurement_y}
\end{equation}
the eigenvectors of $\sigma_y$.
This is coincided with the result shown below equation~(\ref{eq:expectation2}) that measuring the observable $\sigma_y^{\otimes N}$ fails to attain Heisenberg limit for the $\varphi=k\pi/N,\,(k\in\mathbb{Z})$ cases when $N$ is even.

\section{Conclusion\label{sec:conclusion}}
We have addressed the optimization problem of measurements for achieving the ultimate sensitivity determined by the QCRB.
From the propagation of error, we derive the N\&S condition of the optimal observables for single parameter estimate by using the SRUR.
As an application of this condition, we examine the optimal observables for GHZ states to achieve the ultimate sensitivity at the Heisenberg limit.
We consider an experimentally feasible case that the observable operators are restricted to separably acting on the subsystem.
We then find the general expression of the optimal separable observable by applying the N\&S condition, and show that it is exactly equivalent to the parity measurement when applying a $\pi/2$ pulse operation.
However, Jose {\em et al} in \cite{Jose&Shaji2013} gave a contradictory conclusion with respect to ours that separable measurements are impossible to beat the shot-noise limit even for entangled states.
We show that for the GHZ state case, their conclusion is established {\em only} for some particular values of the parameter.
Our results may be helpful for further investigation of the quantum metrology.

\ack
We would like to thank Dr. Heng-Na Xiong and Dr. Qing-Shou Tan for helpful discussions.
We also thank the second referee for constructive suggestions.
This work is supported by the NFRPC with Grant No. 2012CB921602, the NSFC with Grants No. 11025527 and No. 10935010, and National Research Foundation and Ministry of Education, Singapore, with Grant No. WBS: R-710-000-008-271.

\end{document}